\documentclass[a4paper,aps,preprint]{revtex4}
\usepackage{graphicx}

\begin{document}

\title{Numerical and Analytical Approach to the Quantum Dynamics 
of Two Coupled Spins in Bosonic Baths}

\author{Alessandro Sergi}
\email{sergi@ukzn.ac.za}

\affiliation{
School of Physics, University of KwaZulu-Natal, Pietermaritzburg,
Private Bag X01 Scottsville, 3209 Pietermaritzburg, South Africa}

\author{Ilya Sinayskiy} 
\email{ilsinay@gmail.com}
\affiliation{
Quantum Research Group, School of Physics, 
University of KwaZulu-Natal, 
Durban, 4001, South Africa}

\author{Francesco Petruccione}
\email{petruccione@ukzn.ac.za}
\affiliation{Quantum Research Group,
School of Physics and National Institute for Theoretical Physics,
University of KwaZulu-Natal, Durban, 4001, South Africa }

\begin{abstract}
The quantum dynamics
of a spin chain interacting with multiple bosonic baths
is described in a mixed Wigner-Heisenberg representation.
The formalism is illustrated by simulating
the time evolution of the reduced density matrix of 
two coupled spins, where each spin is also
coupled to its own bath of harmonic oscillators.
In order to prove the validity of the approach,
an analytical solution in the Born-Markov approximation
is found.
The agreement between the two methods is shown.
\end{abstract}

\maketitle

\section{Introduction}

For the sake of studying quantum information transport in solid state
devices, the quantum dynamics of spin chains coupled to bosonic baths
has attracted much attention in the recent scientific 
literature~\cite{ilya,quiroga,burgarth,braun,zanardi,storcz,dube,campagnano}.
Here, we show how a mixed Wigner-Heisenberg representation of quantum mechanics
is particularly well-suited to the numerical simulation of such systems.
This is illustrated by studying the time evolution of the reduced density
matrix of a minimal chain, composed of two spins, each coupled to a
bath of harmonic oscillators. The temperature of each bath
can be defined independently, so that nonequilibrium situations can 
be addressed with no further theoretical or computational efforts.
The dynamics of the total systems, spins plus harmonic oscillators,
is unitary and numerically exact. No Markovian 
or rotating waves approximations need to be invoked.
Reduced operators are obtained simply by integrating the coordinates
of the oscillators in Wigner phase space.
Our numerical solution is compared with an analytical solution
of the Markovian master equation of the two spins
and good agreement is found.
It is very easy to extend the algorithm to study longer chains
and multiple bosonic baths.

It is worth remarking that the mixed Wigner-Heisenberg representation
that we adopt in this
paper has been originally proposed for introducing a quantum-classical
representation of systems immersed in gravitational fields
and in plasma physics~\cite{qc}. 
In particular it has been developed~\cite{kc}
and applied to a variety of models in chemical physics~\cite{sergi-rate,gabe},
and it has already been noted~\cite{donal} that such a representation is
exact in the case of (bosonic) bath of harmonic oscillators.

This paper is organized as follows.
Section~\ref{sec:wh} illustrates the Wigner-Heisenberg representation
of quantum mechanics.
Section~\ref{sec:model} provides the details of the 
model we have studied.
The numerical algorithms for the computer simulation is illustrated
in Sec.~\ref{sec:algo}.
The Born-Markov approximation for the master equation
and details of the analytical solution are given in Sec.~\ref{sec:ana}.
Results of both our numerical and analytical studies are
displayed in Sec.~\ref{sec:results}.
Finally, our conclusions are reported in Sec.~\ref{sec:conclusions}.

\section{Wigner-Heisenberg Representation of Quantum Mechanics}
\label{sec:wh}

Let us consider a system defined by the total Hamiltonian operator
\begin{equation}
\hat{H}=\hat{H}_S+\hat{H}_B+\hat{H}_{SB}\;,
\label{eq:tot-ham}
\end{equation}
where the subscripts $S$, $B$, and $SB$ stand for subsystem, bath,
and coupling, respectively.
The Heisenberg equation of motion of the density matrix 
can be written as~\cite{b3}
\begin{eqnarray}
\frac{\partial}{\partial t}\hat{\rho}
&=&-\frac{i}{\hbar}\left[\begin{array}{cc} \hat{H} & 
\hat{\rho}\end{array}\right]
\cdot\mbox{\boldmath$\cal B$}\cdot
\left[\begin{array}{c}\hat{H}\\ \hat{\rho}\end{array}\right]
\;,\label{eq:vonneuman}
\end{eqnarray}
where $\mbox{\boldmath$\cal B$}$ is the antisymmetric constant matrix
\begin{equation}
\mbox{\boldmath$\cal B$}=\left[\begin{array}{cc} 0 & 1 \\ -1 & 0\end{array}\right]
\;.
\end{equation}

Assuming that the bath Hamiltonian depends on a pair of canonically 
conjugated operators $\hat{X}=(\hat{R},\hat{P})$, and the coupling has
the form $\hat{H}_{SB}=\hat{H}_{SB}(\hat{R})$,
we can introduce a partial Wigner transform for the density matrix
\begin{eqnarray}
\hat{\rho}_W(X)
&=&\frac{1}{(2\pi\hbar)^{3N}}
\int dz e^{iP\cdot z/\hbar}
\langle R-\frac{z}{2}\vert\hat{\rho}
\vert R+\frac{z}{2}\rangle \;,
\end{eqnarray}
and for the generic bath-dependent operator 
$\hat{\chi}(\hat{R},\hat{P})$
\begin{eqnarray}
\hat{\chi}_W(X)
&=&
\int dz e^{iP\cdot z/\hbar}
\langle R-\frac{z}{2}\vert\hat{\chi}
\vert R+\frac{z}{2}\rangle \;,
\end{eqnarray}
where $X=(R,P)$ are canonically conjugated classical 
variables in phase space.
Taking the partial Wigner transform of Eq.~(\ref{eq:vonneuman})
\begin{eqnarray}
\frac{\partial}{\partial t}\hat{\rho}_W(X,t)
&=&-\frac{i}{\hbar}\left[\begin{array}{cc} \hat{H}_W(X) & 
\hat{\rho}\end{array}\right]
\cdot\mbox{\boldmath$\cal D$}\cdot
\left[\begin{array}{c}\hat{H}_W(X)\\ \hat{\rho}_W(X,t)\end{array}\right]
\;,\nonumber\\
\label{eq:pW}
\end{eqnarray}
where
\begin{eqnarray}
\mbox{\boldmath$\cal D$}
&=&
\left[\begin{array}{cc} 1 & e^{\frac{i\hbar}{2}
\overleftarrow{\partial}_I{\cal B}_{IJ}
\overrightarrow{\partial}_J}
\\
-e^{\frac{i\hbar}{2}
\overleftarrow{\partial}_I{\cal B}_{IJ}
\overrightarrow{\partial}_J}
& 0\end{array}\right]\;.\label{eq:defD}
\end{eqnarray}
In Equation~(\ref{eq:defD}) we have used the symbol
$\overrightarrow{\partial}_I=\overrightarrow{\partial}/\partial X_I$
to denote an operator of derivation (with respect to the phase space point
coordinates) which acts on whatever stands on its right.
Analogously, $\overleftarrow{\partial}_I$ acts on whatever stands on its left.
Moreover, the summation over repeated indices must be performed in
Eq.~(\ref{eq:defD}) and in the following.
The mixed Wigner-Heisenberg form of the Hamiltonian
operator, $\hat{H}_W$, is 
\begin{equation}
\hat{H}_W(X)=\hat{H}_S+H_{W,B}(X)+\hat{H}_{W,SB}(R)\;.
\end{equation}

Equation~(\ref{eq:pW}) provides a mixed Wigner-Heisenberg
representation of quantum mechanics, where operators also depend
on phase space (c-number) coordinates, which is completely equivalent
to the usual Heisenberg representation.
However, the difficulties associated to the solution of Eq.~(\ref{eq:pW})
are formidable. Yet, for quadratic bath Hamiltonians
\begin{equation}
\hat{H}_{W,B}=\sum_{I=1}^N\left
(\frac{P_I^2}{2}+\frac{1}{2}\omega_I^2R_I^2\right)\;,
\label{eq:hambqua}
\end{equation}
where $(R_I,P_I)$, $I=1,\ldots,N$, are the coordinates and momenta,
respectively,
of a system of $N$ independent harmonic oscillators with frequencies
$\omega_I$,
and for interaction Hamiltonians of the type
\begin{equation}
\hat{H}_{W,SB}=V_B(R)\otimes \hat{H}_{S}^{\prime}\;,
\label{eq:hamcqua}
\end{equation}
where $V_B(R)$ is at most a quadratic function of $R$
and $\hat{H}_S^{\prime}$ acts only in the Hilbert space
of the subsystem,
Eq.~(\ref{eq:pW}) can be rewritten using
the antisymmetric operator matrix
\begin{eqnarray}
\mbox{\boldmath$\cal D$}_{\rm lin}
&=&
\left[\begin{array}{cc} 1 & 1+\frac{i\hbar}{2}
\overleftarrow{\partial}_I{\cal B}_{IJ}
\overrightarrow{\partial}_J
\\
-1-\frac{i\hbar}{2}
\overleftarrow{\partial}_I{\cal B}_{IJ}
\overrightarrow{\partial}_J
& 0\end{array}\right]\;.
\nonumber\\
\end{eqnarray}
Actually, it can be shown that for the class of Hamiltonians
specified by Eqs.~(\ref{eq:hambqua}) and~(\ref{eq:hamcqua})
\begin{equation}
\mbox{\boldmath$\cal D$}\to\mbox{\boldmath$\cal D$}_{\rm lin}
\label{eq:Dlinsub}
\end{equation}
holds exactly.
For more general bath Hamiltonians and couplings, such a substitution
amounts to performing a quantum-classical approximation~\cite{kc}.
What matters here is that for the class of systems we are interested in
the Eq.~(\ref{eq:Dlinsub}) is exact and provides 
via Eq.~(\ref{eq:vonneuman}) a Wigner-Heisenberg formulation of quantum
mechanics which can be numerically simulated employing algorithms
previously developed within a chemical-physical
context~\cite{theorchemacc}.

\section{Model system}\label{sec:model}

The system we are interested in this paper is defined by the following
subsystem Hamiltonian
\begin{equation}
\hat{H}_S=
-j_x\hat{\sigma}_x^{(1)}\hat{\sigma}_x^{(2)}
-j_y\hat{\sigma}_y^{(1)}\hat{\sigma}_y^{(2)}
-j_z\hat{\sigma}_z^{(1)}\hat{\sigma}_z^{(2)}
\label{eq:h1}
\end{equation}
representing a chain of two quantum spins
coupled to each other.
The constants $j_i$, with $i=x,y,z$, dictate
the strength of the coupling between the spins.
The operators $\hat{\sigma}_i^{(k_s)}$
with $i=x,y,z$ are the Pauli matrix operators
for spin $k_s=1,2$.
The bath Hamiltonian is
\begin{equation}
H_{W,B}=
\sum_{k_s=1}^2\sum_{I=1}^N\frac{P_{I,k_s}^2}{2}
+
\frac{\omega_I^2}{2}R_{I,k_s}^2\;.
\label{eq:h2}
\end{equation}
The above Hamiltonian represents two independent harmonic oscillator baths
with coordinates and momenta $(R_{I,k_s},P_{I,k_s})$,
(where $I=1,N$ labels the oscillators and $k_s=1,2$
labels the bath). 
The harmonic oscillator frequencies $\omega_I$
are taken to be bath-independent since
we want to adopt two baths
with identical spectral density.
However, the baths can have different initial
conditions (and eventually different temperature).
The coupling is given by
\begin{equation}
\hat{H}_{W,SB}=-\sum_{k_s=1}^2\sum_{I=1}^Nc_IR_{I,k_s}\hat{\sigma}_z^{(k_s)}
\;,\label{eq:h3}
\end{equation}
showing that each spin is coupled to its own oscillator bath.

The density matrix of the two-spin chains obeys the exact 
Wigner-Heisenberg equation
\begin{eqnarray}
\frac{\partial}{\partial t}\hat{\rho}_W
&=&-\frac{i}{\hbar}\left[\begin{array}{cc} \hat{H}_W & 
\hat{\rho}_W\end{array}\right]
\cdot\mbox{\boldmath$\cal D$}_{\rm lin}\cdot
\left[\begin{array}{c}\hat{H}_W\\ \hat{\rho}_W\end{array}\right]
\;,\label{eq:vonneuman-lin}
\end{eqnarray}
where $\hat{H}_W=\hat{H}_S+H_{W,B}+\hat{H}_{W,SB}$
is given by the sum of Eqs.~(\ref{eq:h1}-\ref{eq:h3}).
The reduced density matrix of the spin subsystem is given at all times
by
\begin{equation}
\hat{\rho}_S(t)=\int\prod_{k_s=1}^2\prod_{I=1}^NdX_{I,k_s}
\hat{\rho}_W(X,t)\;.
\end{equation}

For the calculation presented in this paper, we assume an initially
uncorrelated density matrix, which, once partially Wigner transformed,
takes the form
\begin{equation}
\hat{\rho}_W(t_0)
=\hat{\rho}_S(t_0)\rho_{W,B}(X,t_0) \;,
\end{equation}
where 
\begin{eqnarray}
\rho_{W,b}(X,t_0)
&=&\prod_{k_s=1}^2\prod_{I=1}^N
\frac{\tanh(\beta_{k_s}\omega_I/2)}{\pi}\nonumber\\
&\times&\exp
\left[-2\frac{\tanh(\beta_{k_s}\omega_I/2)}{\omega_I}
H_{W,B}\right]\;,
\label{eq:rhoWb}
\end{eqnarray}
and where $H_{W,B}$ is defined in Eq.~(\ref{eq:h2})
and $\beta_{js}=(k_BT_{k_s})^{-1}$ is the inverse temperature
of each oscillator bath ($k_B$ is the Boltzmann constant).

\section{Numerical Algorithm}\label{sec:algo}

In cases in which the coupling Hamiltonian $\hat{H}_{W,SB}$
can be treated as a small perturbation (weak coupling),
it is useful to represent the abstract Eq.~(\ref{eq:vonneuman-lin})
in the adiabatic basis. Such a basis is defined
by the eigenvalue equation 
\begin{equation}
(\hat{H}_S+\hat{H}_{W,SB})|\alpha;R\rangle
=E_{\alpha}(R)|\alpha;R\rangle \;.\label{eq:adbasis}
\end{equation}
Hence, Eq.~(\ref{eq:vonneuman-lin}) can be recast in propagator form
\begin{equation}
\rho_W^{\alpha\alpha'}(X,t)
=\sum_{\beta\beta'}
\left(e^{-it{\cal L}}\right)_{\alpha\alpha',\beta\beta'}
\rho_W^{\beta\beta'}(X)\;,
\end{equation}
where 
\begin{eqnarray}
i{\cal L}_{\alpha\alpha',\beta\beta'}
&=&i{\cal L}^0_{\alpha\alpha'}
\delta_{\alpha\alpha'}\delta_{\beta\beta'}
+{\cal T}_{\alpha\alpha',\beta\beta'}
\;.
\end{eqnarray}
The operator $i{\cal L}_{\alpha\alpha'}^0$ is defined as
\begin{equation}
i{\cal L}_{\alpha\alpha'}^0=
i\omega_{\alpha\alpha'}+iL_{\alpha\alpha'}\;,
\end{equation}
where $\omega_{\alpha\alpha'}=(E_{\alpha}(R)-E_{\alpha'}(R))/\hbar$
and
\begin{eqnarray}
iL_{\alpha\alpha'}=P\frac{\partial}{\partial R}
+\frac{1}{2}(F_W^{\alpha}+F_W^{\alpha'})\cdot
\frac{\partial}{\partial P}
\;.\label{eq:Lalphaalpha'}
\end{eqnarray}
$F_W^{\alpha}=-\langle\alpha;R|\partial\hat{H}_W/\partial R
|\alpha;R\rangle$ is the Hellmann-Feynman force~\cite{hf-force}.
The transition operator ${\cal T}_{\alpha\alpha',\beta\beta'}$
is purely off-diagonal
and defined by
\begin{eqnarray}
{\cal T}_{\alpha\alpha',\beta\beta'}
&=&
P\cdot d_{\alpha\beta}
\left(1+\frac{1}{2}\frac{(E_{\alpha}-E_{\beta})d_{\alpha\beta}}
{P\cdot d_{\alpha\beta}}\cdot
\frac{\partial}{\partial P}\right)
\delta_{\alpha'\beta'}
\nonumber\\
&+&P\cdot d^*_{\alpha'\beta'}
\left(1+\frac{1}{2}\frac{(E_{\alpha'}-E_{\beta'})d^*_{\alpha'\beta'}}
{P\cdot d^*_{\alpha'\beta'}}\cdot
\frac{\partial}{\partial P}\right)
\delta_{\alpha\beta}\nonumber\\
\end{eqnarray}
where 
$d_{\alpha\beta}=\langle\alpha;R|
\overrightarrow{\partial}/\partial R|\beta;R\rangle$
is the coupling vector between the adiabatic states
$|\alpha;R\rangle$ defined in Eq.~(\ref{eq:adbasis}).
Above and in the following, the quantities are defined adopting
scaled coordinates, according to the definition
of the Hamiltonians in Eqs.~(\ref{eq:h1}-\ref{eq:h3}).
The operator ${\cal T}_{\alpha\alpha',\beta\beta'}$
realises the quantum transitions of the subsystem due to the
coupling to the bath. Assuming weak coupling, and for the sake of
comparison to a Markovian master equation, the action of the transition
operator will be disregarded: This amounts to perform an adiabatic
approximation of the dynamical evolution of the spin subsystem. 

In the adiabatic approximation the evolution of the density matrix
becomes simply
\begin{eqnarray}
\rho_W^{\alpha\alpha'}(X,t)
&=&
e^{-it{\cal L}^0_{\alpha\alpha'}}
\rho_W^{\alpha\alpha'}(X)
\nonumber\\
&=&
e^{-i\int_0^td\tau\omega_{\alpha\alpha'}}
e^{-itL_{\alpha\alpha'}}
\rho_W^{\alpha\alpha'}(X)\;.
\label{eq:adiabatic-prop}
\end{eqnarray}
Equation~(\ref{eq:adiabatic-prop}) shows that using the adiabatic
approximation, in the adiabatic basis, the evolution of the density
matrix in the Wigner-Heisenberg representation can be calculated
by propagating classical-like trajectories,
under the action of the Liouville operator~(\ref{eq:Lalphaalpha'}),
and considering a phase factor integrated along the trajectory.
The initial $X$ coordinates, representing the quantum state of
the bath in phase space, can be sampled from 
the initial density matrix~(\ref{eq:rhoWb}).

\begin{figure}
\includegraphics{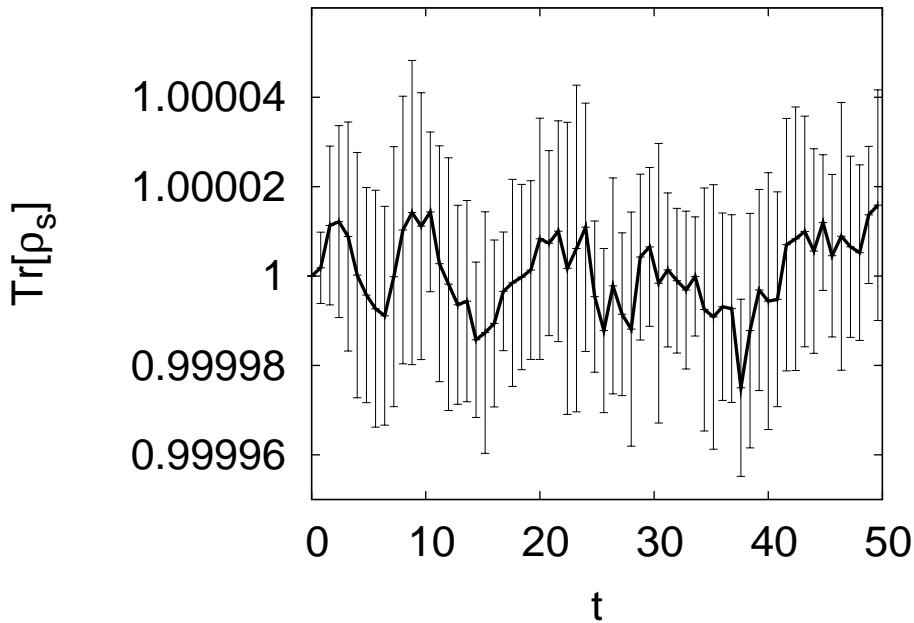}
\caption{
Time evolution trace of  the reduced density matrix element
vs time ($\beta_1=\beta_2=0.005$).
Initial density matrix $\hat{\rho}_s(0)=|1,0\rangle\langle 1,0|$.
The error bars display the numerical error.
}
\label{fig:fig1}
\end{figure}

Although in the adiabatic approximation the dynamics
is easily calculated in the adiabatic basis,
for quantum information problems
it is more convenient to consider the reduced
density matrix
\begin{equation}
\rho_W^{\mu\nu}(t)
=\int dX \sum_{\alpha\alpha'}U_{\mu\alpha}(R)
\rho_W^{\alpha\alpha'}(X,t)
({\bf U}^{-1})_{\alpha'\nu}(R)
\label{eq:redrhodia}
\end{equation}
in the natural basis $|1\rangle=|1,1\rangle$, $|2\rangle=|1,0\rangle$,
$|3\rangle=|0,1\rangle$, $|4\rangle=|0,0\rangle$.
The matrix $\bf U$ appearing in Eq.~(\ref{eq:redrhodia})
is, of course, the rotation matrix from the adiabatic to
the natural basis which can be constructed, as well known,
by using the adiabatic eigenvectors as columns.

Everything seems quite straightforward so far. However, the definition
of $\bf U$ is somewhat arbitrary, since the columns can be
evenly  permuted, and the adiabatic eigenvectors 
in the Wigner-Heisenberg representation of quantum mechanics depend
on the configuration point $R$.
In addition, the LAPACK~\cite{lapack} numerical routines, which
we have used to calculate
the eigenvectors, return a matrix $\bf U$ with the columns ordered
corresponding to the increasing value of the eigenvalues.
It turns out that this configuration-dependent permutation
of the columns of $\bf U$ introduces fictitious dynamics,
as can be verified by propagating the density matrix
$\hat{\rho}=|1\rangle\langle 1|$, defined in terms of the natural
state ``spin-up spin-up'' of the spin chain,
which should be left invariant under the action of the
Hamiltonian $\hat{H}_S+\hat{H}_{W,SB}$, defined
in Eqs.~(\ref{eq:h1}) and~(\ref{eq:h3}).

In order to solve this problem, it is sufficient to note that
one would like to have a rotation matrix $\bf U$ as close
as it could be to the matrix $\mbox{\boldmath$\cal E$}$
formed by ordering the Cartesian basis vectors ${\bf e}^j$
(in the present case $j=1,\ldots,4$), with
${\bf e}^1=[\begin{array}{cccc} 1 & 0 & 0 & 0\end{array}]$,
${\bf e}^2=[\begin{array}{cccc} 0 & 1 & 0 & 0\end{array}]$
and so on.
Upon writing ${\bf u}^{\alpha}$ for the adiabatic eigenvectors,
$\alpha=1,\ldots, 4$, one can define a metric 
\begin{equation}
g^{\alpha,j}=({\bf u}^{\alpha}-{\bf e}^j)\cdot
({\bf u}^{\alpha}-{\bf e}^j) \;.\label{eq:g}
\end{equation}
The definition of the metric in Eq.~(\ref{eq:g})
allows us to solve the ordering problem in a unique way.
As a matter of fact, for each $j$, labelling the columns of the desired
rotation matrix, we can look for the $\alpha$ which minimizes
the metric $g^{\alpha,j}$: This leads to the possibility
of ordering the columns of $\bf U$ in such a way that
this matrix is as \emph{close} as it can be to 
$\mbox{\boldmath$\cal E$}$, and it effectively solves
the numerical problem with the fictitious dynamics
arising from the permutations of the adiabatic eigenvectors
along the phase space trajectory.

\begin{figure}
\includegraphics{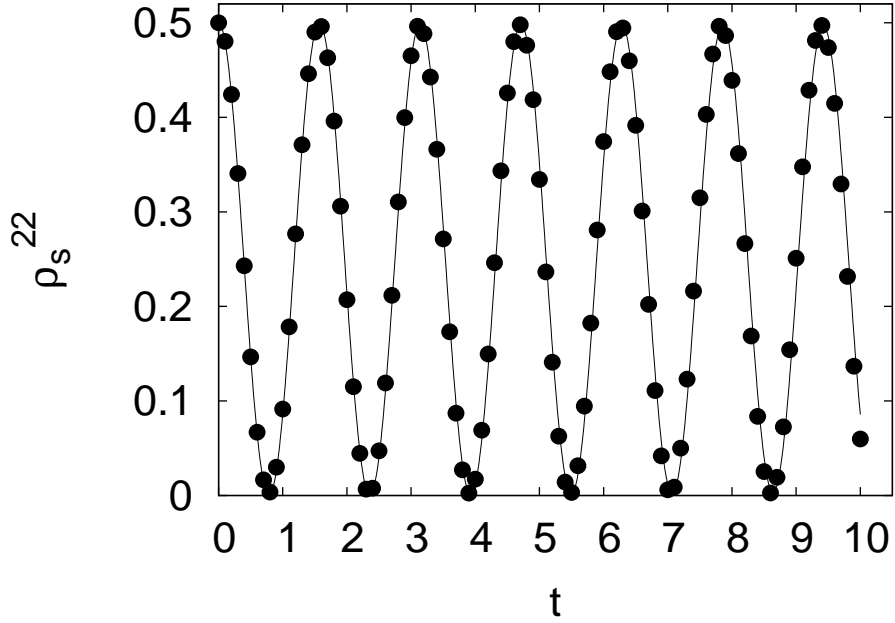}
\caption{
Time evolution of the reduced density matrix element
$\rho_s^{22}$ vs time ($\beta_1=1$, $\beta_2=0.3$). 
Initial density matrix
 $\hat{rho_s}(0)=|\Psi_0\rangle\langle\Psi_0|$
with $|\Psi_0\rangle=(|1,1\rangle-|1,0\rangle)/sqrt{2}$.
The continuous line is the analytical
solution. The filled circles display the results of the numerical
calculation.
}
\label{fig:fig2}
\end{figure}

\begin{figure}
\includegraphics{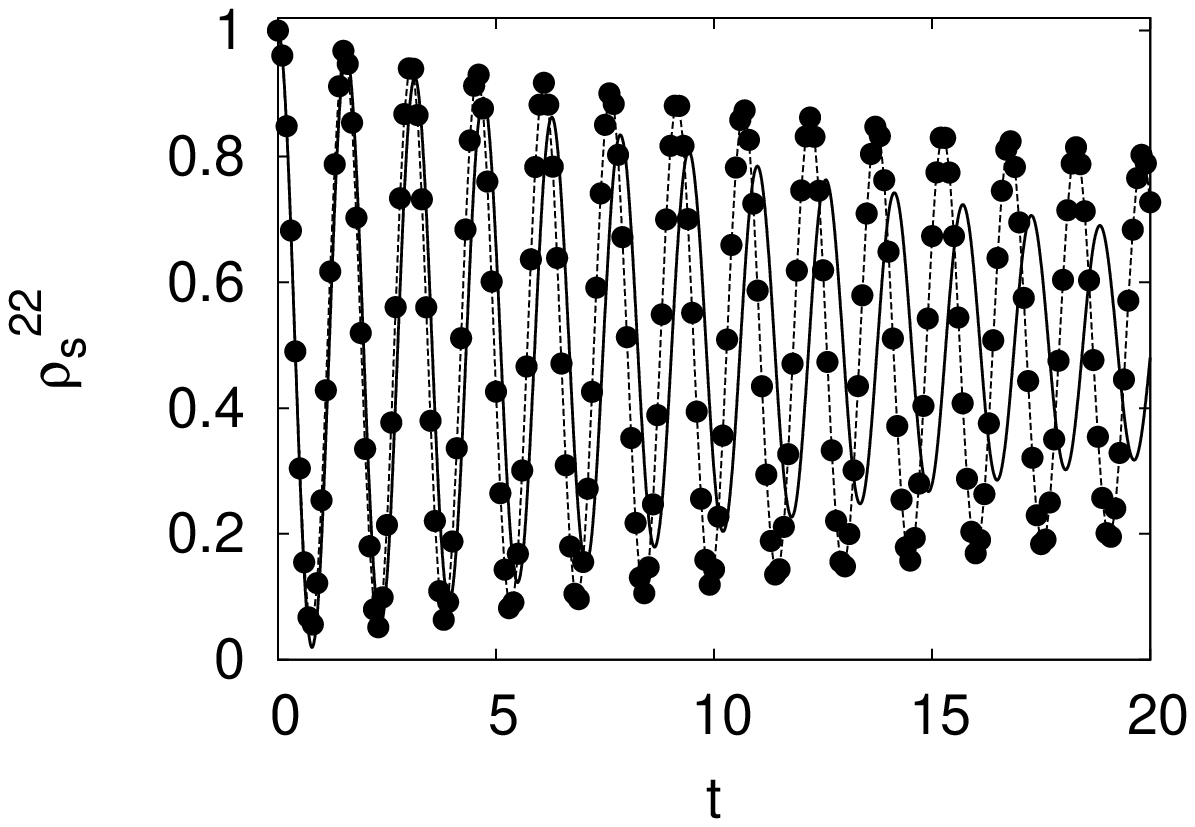}
\caption{
Time evolution of the reduced density matrix element
$\rho_s^{22}$ vs time $(\beta_1=\beta_2=0.005)$.
Initial density matrix $\hat{\rho}_s(0)=|1,0\rangle\langle 1,0|$.
 The continuous line is the analytical
solution. The filled circles display the results of the numerical
calculation. They are joined by a dashed line to help the eye.
}
\label{fig:fig3}
\end{figure}

\section{Master equation for the coupled spins}\label{sec:ana}

A system with total Hamiltonian~(\ref{eq:tot-ham}),
obeying the Liouville (Heisenberg) 
equation of motion~(\ref{eq:vonneuman}),
can be studied in the weak coupling limit
by performing the Born-Markov approximation~\cite{toqs}.
In such a case, the equation
for the reduced density matrix becomes
\begin{eqnarray}
& &\frac{d}{dt}\hat{\rho}_S^{(I)}(t)= \nonumber\\
& &-\int_0^\infty
ds\mathrm{tr}_B[\hat{H}_{SB}^{(I)}(t),
[\hat{H}_{SB}^{(I)}(t-s),\hat{\rho}_S^{(I)}(t)\otimes\hat{\rho}_B(0)]],
\end{eqnarray}
where the index $I$ denotes the interaction picture with respect
to the free Hamiltonians of the system and bath. The operator
$\hat{\rho}_S$ denotes the reduced density matrix of the system
$S$ and $\hat{\rho}_B$ is the density matrix of the reservoir $B$.

After performing the rotating wave approximation over the rapidly
oscillating term in the master equation one gets:
\begin{eqnarray}
\frac{d}{dt}\hat{\rho}_S(t)
&=&-i[\hat{H}_S,\hat{\rho}_S(t)]+\label{mseq}\nonumber\\
& &
\sum_\omega\sum_{\alpha,\beta}
\gamma_{\alpha,\beta}(\omega)
\Big(\hat{V}_\beta(\omega)\hat{\rho}_S(t)\hat{V}_\alpha^\dag(\omega)-
\nonumber\\
& &\qquad
\frac{1}{2}\left[\hat{V}_\alpha^\dag(\omega)
\hat{V}_\beta(\omega),\hat{\rho}_S(t)\right]_+\Big) .
\end{eqnarray}

To obtain Eq.~(\ref{mseq}) one assumes that the
system-environment interaction has the form $\hat{H}_{SB}=\sum_i
\hat{V}_i\otimes\hat{f}_i$; the operators
$\hat{V}_i=\hat{V}_i^\dag$ and $\hat{f}_i=\hat{f}_i^\dag$ act on
the system and the bath degrees of freedom, respectively. In Eq.
(\ref{mseq}) a \emph{Lamb-type} renormalization Hamiltonian was
neglected and decay rates $\gamma_{\alpha,\beta}(\omega)$ are
given by the Fourier image of the bath correlation functions:
\begin{equation}
\gamma_{\alpha,\beta}(\omega)=\int_{-\infty}^{+\infty}ds
e^{i\omega
s}\langle\hat{f}_{\alpha}^\dag(s)\hat{f}_{\beta}(0)\rangle.
\end{equation}

The transition operators $\hat{V}_\alpha(\omega)$ originates from
the decomposition of the operator $\hat{V}_\alpha$ in the basis of
the eigenoperators of the system Hamiltonian $\hat{H}_S$. If one
denotes the eigenvalues of the Hamiltonian $\hat{H}_S$ by
$\varepsilon$ and the corresponding projection operator as
$\hat{\Pi}(\varepsilon)$ then:
\begin{equation}
\hat{V}_\alpha(\omega)=\sum_{\varepsilon'-\varepsilon=\omega}\hat{\Pi}(\varepsilon)\hat{V}_{\alpha}\hat{\Pi}(\varepsilon').
\end{equation}

To obtain the master equation for the open system we rewrite the
Hamiltonian of the whole system in the following way:
\begin{equation}
\hat{H}=\hat{H}_{S}+\hat{H}_{B1}+\hat{H}_{B2}+\hat{H}_{SB1}+\hat{H}_{SB2},\label{ham}\end{equation}
where 
$\hat{H}_{S}$ is defined in Eq.~(\ref{eq:h1})
and here we further assume that $j_x=j_y=j$,
so that 
the constants $j\geq0$ and $j_z\geq0$ denote the strenght of XY
and ZZ interaction, respectively.
As already stated in the previous section, in this article
scaled units are chosen, so that $k_B=\hbar=1$.
We rewrite the Hamiltonians of the
reservoirs $k_s=1,2$ as
\begin{equation}
\hat{H}_{Bk_s}=\sum_n \omega_{n,k_s}\hat{b}^\dag_{n,k_s}\hat{b}_{n,k_s}.
\end{equation}
The interaction between the spin subsystem and the 
bosonic baths is described by
\begin{equation}
\hat{H}_{SBk_s}=
-\hat{\sigma}^{(k_s)}_z\sum_{n}g_{n}^{(k_s)}
(\hat{b}_{n,k_s}+\hat{b}_{n,k_s}^{\dag}).
\end{equation}

To derive an equation of the form (\ref{mseq}) for the
Hamiltonian (\ref{ham}) one needs to find the eigenvalues and
eigenvectors of the Hamiltonian $\hat{H}_S$ (\ref{eq:h1}):
\begin{eqnarray}
\hat{H}_S&=&\sum_{i=1}^4
\lambda_i|\lambda_i\rangle\langle\lambda_i|\;, 
\end{eqnarray}
namely
\begin{eqnarray}
|\lambda_1\rangle&=&|1,1\rangle, \quad \lambda_1=-j_z\;,\\
|\lambda_2\rangle&=&|0,0\rangle, \quad \lambda_2=-j_z\;,\\
|\lambda_3\rangle&=&
\frac{1}{\sqrt{2}}\left(|1,0\rangle+|0,1\rangle\right),
\quad \lambda_3=-2j+j_z\;, \\
|\lambda_4\rangle&=&
\frac{1}{\sqrt{2}}\left(-|1,0\rangle+|0,1\rangle\right), 
\quad \lambda_4=2j+j_z\;.
\end{eqnarray}
In this basis, the transition operators take the form:
\begin{equation}
V_0^{(1)}=V_0^{(2)}=|\lambda_1\rangle\langle
\lambda_1|-|\lambda_2\rangle\langle\lambda_2|,
\end{equation}
with $\omega_0=0$. The operators $V_0$ cause decoherence.
The operators $V$
\begin{eqnarray}
V_{(1)}&=&-|\lambda_3\rangle\langle\lambda_4|,
\\
V_{(2)}&=&|\lambda_3\rangle\langle\lambda_4|,
\end{eqnarray}
describe the dissipation between the levels
$\lambda_3$ and $\lambda_4$ with
the transition frequency $\omega=4j$.
Finally, the master equation takes the form:
\begin{equation}
\frac{d\hat{\rho}}{dt}=
-i[\hat{H}_{S},\hat{\rho}]+\sum_{i=1}^2\left({\cal
L}_{Di}(\hat{\rho})+{\cal L}_{Ci}(\hat{\rho})\right),
\label{eq:lind}
\end{equation}
where
\begin{eqnarray}
{\cal
L}_{Di}(\hat{\rho})&=&\gamma^{(i)}(-\omega)\left(\hat{V}_{(i)}\hat{\rho}\hat{V}_{(i)}^\dag-\frac{1}{2}\left[\hat{V}_{(i)}^\dag\hat{V}_{(i)},\hat{\rho}\right]_+\right)+\\
& &
\gamma^{(i)}(\omega)\left(\hat{V}_{(i)}^\dag\hat{\rho}\hat{V}_{(i)}-\frac{1}{2}\left[\hat{V}_{(i)}\hat{V}_{(i)}^\dag,\hat{\rho}\right]_+\right)
\end{eqnarray}
and
\begin{eqnarray}
{\cal
L}_{Ci}(\hat{\rho})&=&\left(\gamma^{(i)}(0_+)+\gamma^{(i)}(0_-)\right)\times\\
& &
\left(\hat{V}_0^{(i)}\hat{\rho}\hat{V}_0^{(i)}-\frac{1}{2}\left[\hat{V}_0^{(i)}\hat{V}_0^{(i)},\hat{\rho}\right]_+\right).
\end{eqnarray}

In the basis of eigenvectors of the Hamiltonian $\hat{H}_S$ the
system of the corresponding differential equations can be solved.
The exact solution of Eq.~(\ref{eq:lind}) is
\begin{equation}
\hat{\rho}(t)=\sum_{i,j=1}^4f_{i,j}(t)|\lambda_i\rangle\langle\lambda_j|,
\end{equation}
where we have introduced the elements of the density matrix:
\begin{equation}
f_{11}(t)=f_{11}(0),
\end{equation}
\begin{equation}
f_{22}(t)=f_{22}(0),
\end{equation}
\begin{eqnarray}
f_{33}(t)&=&\left(\Omega_-+\Omega_+e^{-\Omega
t}\right)\frac{f_{33}(0)}{\Omega}+\\\nonumber &
&\left(1-e^{-\Omega t}\right)\Omega_-\frac{f_{44}(0)}{\Omega},
\end{eqnarray}
\begin{eqnarray}
f_{44}(t)&=&\left(1-e^{-\Omega
t}\right)\Omega_-\frac{f_{33}(0)}{\Omega}+\\\nonumber & &
\left(\Omega_++\Omega_-e^{-\Omega
t}\right)\frac{f_{44}(0)}{\Omega},
\end{eqnarray}
\begin{equation}
f_{12}(t)=f_{12}(0)\exp{\left(-4g_ct+it
(\lambda_2-\lambda_1)\right)},
\end{equation}

\begin{equation}
f_{13}(t)=f_{13}(0)\exp{\left(-g_ct-\Omega_+t+it
(\lambda_3-\lambda_1)\right)}, \end{equation}

\begin{equation}
f_{14}(t)=f_{14}(0)\exp{\left(-g_ct-\Omega_-t+it
(\lambda_4-\lambda_1)\right)}, \end{equation}

\begin{equation}
f_{23}(t)=f_{23}(0)\exp{\left(-g_ct-\Omega_+t+it
(\lambda_3-\lambda_2)\right)}, \end{equation}

\begin{equation}
f_{24}(t)=f_{24}(0)\exp{\left(-g_ct-\Omega_-t+it
(\lambda_4-\lambda_2)\right)}, \end{equation}

\begin{equation}
f_{34}(t)=f_{34}(0)\exp{\left(-\Omega t+it
(\lambda_4-\lambda_3)\right)}.
\end{equation}

In the above expressions we have defined the constants
\begin{equation}
\Omega_\pm=\frac{1}{2}\left(\gamma^{(1)}(\pm\omega)+\gamma^{(2)}(\pm\omega)\right),
\end{equation}
\begin{equation}
\Omega=\Omega_++\Omega_-,
\end{equation}
\begin{equation}
g_c=\frac{1}{2}\sum_{i=1}^2\left(\gamma^{(i)}(0_+)
+\gamma^{(i)}(0_-)\right)\;.
\end{equation}
The above solution will be used in the
following as a reference for the numerical simulation.


\section{Calculations and Results} \label{sec:results}

We have performed various numerical calculations varying the
temperatures of the oscillator baths
and compared to the analytical solution given in Sec.~\ref{sec:ana}.
The coupling constants in the Hamiltonian~(\ref{eq:h1})
have been taken as $j_x=j_y=j=1$ and $j_z=1/2$.
The two baths, with $N=200$ harmonic oscillators each,
have been assigned an Ohmic spectral density.
To this end we employed the form of the coupling constants $c_I$
and frequencies $\omega_I$ introduced in Ref.~\cite{makri}:
\begin{eqnarray}
c_I&=&\left(\xi\omega_0\omega_j\right)^{1/2}\\
\omega_I&=&-\ln\left(1-I\omega_0\right)
\end{eqnarray}
where
$\omega_0=(1-\exp(-\omega_{\rm max})/N$, with $\xi=0.007$
and $\omega_{\rm max}=3$.
In order to compare with the analytical solutions
of the weak-coupling master equation of Sec.~\ref{sec:ana},
we have performed an adiabatic propagation in the mixed
Wigner-Heisenberg representation of quantum mechanics
and sampled 50000 initial conditions to calculate
the reduced density matrix, $\hat{\rho}_S$ of the two coupled spins.
Figure~\ref{fig:fig1} shows the numerical precision
of our numerical scheme displaying the constancy
of the trace of $\hat{\rho}_S$ versus time in the case of
$\beta_1=\beta_2=0.005$.

In general, we have found a very good agreement between
the results provided by both the numerical and the analytical
approach for all the various temperatures investigated.
Here, we discuss explicitly two calculations.

Calculation (i) has been performed with the baths
in a nonequilibrium configuration, at the two different
temperatures $\beta_1=0.3$  and $\beta_2=1$.
The initial reduced density matrix $\hat{\rho}_S(0)$
has been taken equal to
$\hat{\rho}_S(0)=|\Psi_0\rangle\langle\Psi_0|$,
with $|\Psi_0\rangle=\frac{1}{\sqrt{2}}(|1,1\rangle-|1,0\rangle)$.
Figure~\ref{fig:fig2} shows the comparison between the numerical
and the analytical dynamics of the matrix element $\rho_S^{22}$.
In this case, the analytical solution is
\begin{equation}
\rho^{22}_S(t)=
\frac{1}{4}\left[1+\exp(-\Omega(\beta_1,\beta_2,\omega) t)
\cos(\omega t)\right]\;.
\end{equation}
Of course, in such a low-temperature case
the Markovian approximation is expected to provide very good results
and this is numerically confirmed.

Calculation (ii) has been performed with
$\beta_1=\beta_2=0.005$ and an initial $\hat{\rho}_s$
equal to $\hat{\rho}_s=|1,0\rangle\langle 1,0|$.
Figure~\ref{fig:fig3} shows the comparison between the numerical
and the analytical dynamics of the matrix element $\rho_S^{22}$.
The theoretical solution is in this case
\begin{equation}
\rho^{22}_s(t)=\frac{1}{2}\left[1+
\exp(-\Omega(\beta_1,\beta_2,\omega) t)\cos(\omega t)\right]\;.
\end{equation}
At higher bath temperature, 
the Markovian approximation (used in the analytical
solution) can describe the numerical results in
a good but qualitative way.
The difference in the oscillation frequencies of
the analytical and the numerical solutions arises
from neglecting the Lamb-type renormalization of the Hamiltonian
$\hat{H}_S$ in the
derivation of the master equation in the Born-Markov approximation.
The discrepancy in the long time decay of the analytical
and numerical results 
arises from the fact that $\Omega(\beta_1,\beta_2,\omega)$ 
in the analytical expression
of $\rho_s^{22}(t)$  should contain some memory effects 
on the time-interval
on which the evolution is considered.


\section{Conclusions} \label{sec:conclusions}

Upon adopting a mixed Wigner-Heisenberg representation,
we have shown how the quantum dynamics
of two coupled spins interacting with multiple bosonic baths
can be numerically simulated.
An analytical solution in the Born-Markov approximation
has also been found and
we have shown agreement between these two approaches.

Both the analytical and the numerical method can be generalized
in order to study additional coupled spins, in order to build longer
spin chains immersed in independent bosonic baths.
Equilibrium and nonequilibrium situation can be addressed on an
equal basis.

The numerical algorithm is suited to include nonadiabatic
correction in the unitary evolution of the density matrix
of the total systems. As such, it can also be used 
to assess novel approaches to non-Markovian dynamics
of open quantum systems.

\section*{Acknowledgments}
This work is based upon research supported by the South
African Research Chair Initiative of the Department of 
Science and Technology and National Research Foundation.


\end{document}